\UseRawInputEncoding
\documentclass[aps,prl,twocolumn,showkeys,groupedaddress,nofootinbib]{revtex4}
\usepackage[pdftex]{graphicx}
\begin{document}
\title{3D cavity-based graphene superconducting quantum circuits in two-qubit architectures}

\author{Kuei-Lin Chiu$^{1,\ddagger,*}$, Avishma J. Lasrado$^{1,\ddagger}$, Cheng-Han Lo$^{1,\ddagger}$, Yen-Chih Chen$^{1}$, Shih-Po Shih$^{1}$, Yen-Hsiang Lin$^{2}$, Chung-Ting Ke$^{3}$}

\affiliation{$^1$Department of Physics, National Sun Yat-Sen University, Kaohsiung 80424, Taiwan}
\affiliation{$^2$Department of physics, National Tsing-Hua University, Hsinchu 300044, Taiwan}
\affiliation{$^3$Institute of Physics, Academia Sinica, Taipei 115201, Taiwan}

\date{\today}

\begin{abstract}

We construct a series of graphene-based superconducting quantum circuits and integrate them into 3D cavities. For a single-qubit device, we demonstrate flux-tunable qubit transition, with a measured $T_1$ $\approx$ 48 ns and a lower bound estimate of $T_2^\ast$ $\approx$ 17.63 ns. By coupling the device to cavities with different resonant frequencies, we access multiple qubit-cavity coupling regimes, enabling the observation of vacuum Rabi splitting and flux-dependent spectral linewidths. In a two-qubit device consisting of a SQUID and a single junction, power-dependent measurements reveal a two-stage dispersive shift. By flux-tuning the cavity frequency at different readout powers, we attribute the first shift to the fixed-qubit and the second to the SQUID-qubit, indicating successful coupling between the two circuits and a single cavity mode. Our study demonstrates the flexible coupling achievable between 2D-material-based superconducting circuits and 3D cavities, and paves the way toward constructing multi-qubit 3D transmon devices from 2D materials.

\end{abstract}

\maketitle

\def\thefootnote{$\ddagger$}\footnotetext{These authors contributed equally to this work}
\def\thefootnote{*}\footnotetext{Corresponding author: K. L. Chiu (klc@mail.nsysu.edu.tw)}
\def\thefootnote{\arabic{footnote}}

\section{I. Introduction}

Integrating low$-$dimensional materials into circuit quantum electrodynamics (cQED) devices has attracted increasing interest due to their wide range of potential applications \cite{Kroll2018,Wang2019,LucasCasparis2018,AlbertHertel2022,Lo2023,Larsen2015,Lange2015,Huo2023,Chiu2017,Schmidt2018,Liu2019,Chiu2020b,Siddiqi2021,Antony2021,Wang2022a,Tanaka2025,Chiu2025a}. In particular, substantial efforts have focused on replacing the insulating tunnel barrier in Josephson junctions (JJs) with low-dimensional materials, enabling the resulting superconducting qubits to inherit their unique properties \cite{Kroll2018,Wang2019,LucasCasparis2018,Larsen2015,Liu2019,Chiu2020b}. Early studies have demonstrated the successful implementation of InAs nanowires, two-dimensional electron gases (2DEGs), and graphene in gate-tunable transmons (gatemons) \cite{Larsen2015,Lange2015,LucasCasparis2018,Kroll2018,Wang2019}, flux-tunable transmons \cite{Lange2015,Luthi2018}, as well as transmons tunable by both gate and flux \cite{Larsen2020}. More recently, weak links based on InAs nanowires and InAs 2DEGs have been integrated into gate-tunable fluxoniums, enabling smooth tuning between heavy and light fluxonium states \cite{Pita-Vidal2020, Strickland2025}. The most recent advances further report transmons based on NbSe$_2$/WSe$_2$/NbSe$_2$ vertical JJs \cite{Balgley2025a}, graphene charge qubits \cite{Aparicio2025}, and weakly anharmonic NbSe$_2$ qubits \cite{DElia2025}. While most studies follow the mainstream approach of superconducting qubits based on 2D coplanar waveguides, which offer advantages for scalability, 3D transmons provide several benefits that their 2D counterparts cannot. First, the external quality factor, cavity decay rate $\kappa$, and qubit–cavity coupling strength $g$ can be conveniently tuned by adjusting the readout pin length and the qubit’s position inside the cavity \cite{Zoepfl2017}. Second, by loading qubits into cavities with different resonant frequencies, the detuning between the qubit frequency maximum and the cavity resonance can be varied, thereby enabling access to different coupling regimes. Third, the use of a 3D cavity simplifies the fabrication process, which in turn improves the yield of devices incorporating new materials. The first two are not achievable with a 2D transmon once its planar readout resonators, capacitors and JJs have been fabricated. In addition, the capacitor pads in 3D transmon can be used as a bond pad in DC transport, allowing comparison between DC and microwave measurements on the same device \cite{Chiu2025}.

In this work, we present a series of graphene-based superconducting circuits in a two-qubit layout coupled to 3D cavities. By loading the same devices into different cavities, thereby varying the detuning between the maximum qubit frequency and the cavity resonance, we access distinct coupling regimes across multiple cooldowns. We construct a graphene single-qubit device with a flux-tunable qubit transition, demonstrating vacuum Rabi splitting in the strong-coupling regime and flux-dependent spectral linewidths in the dispersive regime. In another device comprising a superconducting quantum interference device (SQUID)-based circuit and a single-junction circuit, we observe distinct two-stage power-dependent dispersive shifts when coupling them to two different cavities. Flux tuning of the cavity responses and the onset power of the dispersive shifts further reveal that the first dispersive shift originates from the single-junction circuit, while the second is associated with the SQUID circuit. Our results highlight the potential of 2D materials for realizing multi-qubit 3D transmon architectures, enabling the possibility of joint readout and the study of interactions between distinct 2D-material-based superconducting circuits.

\begin{figure}[!t]	
\includegraphics[scale=0.4]{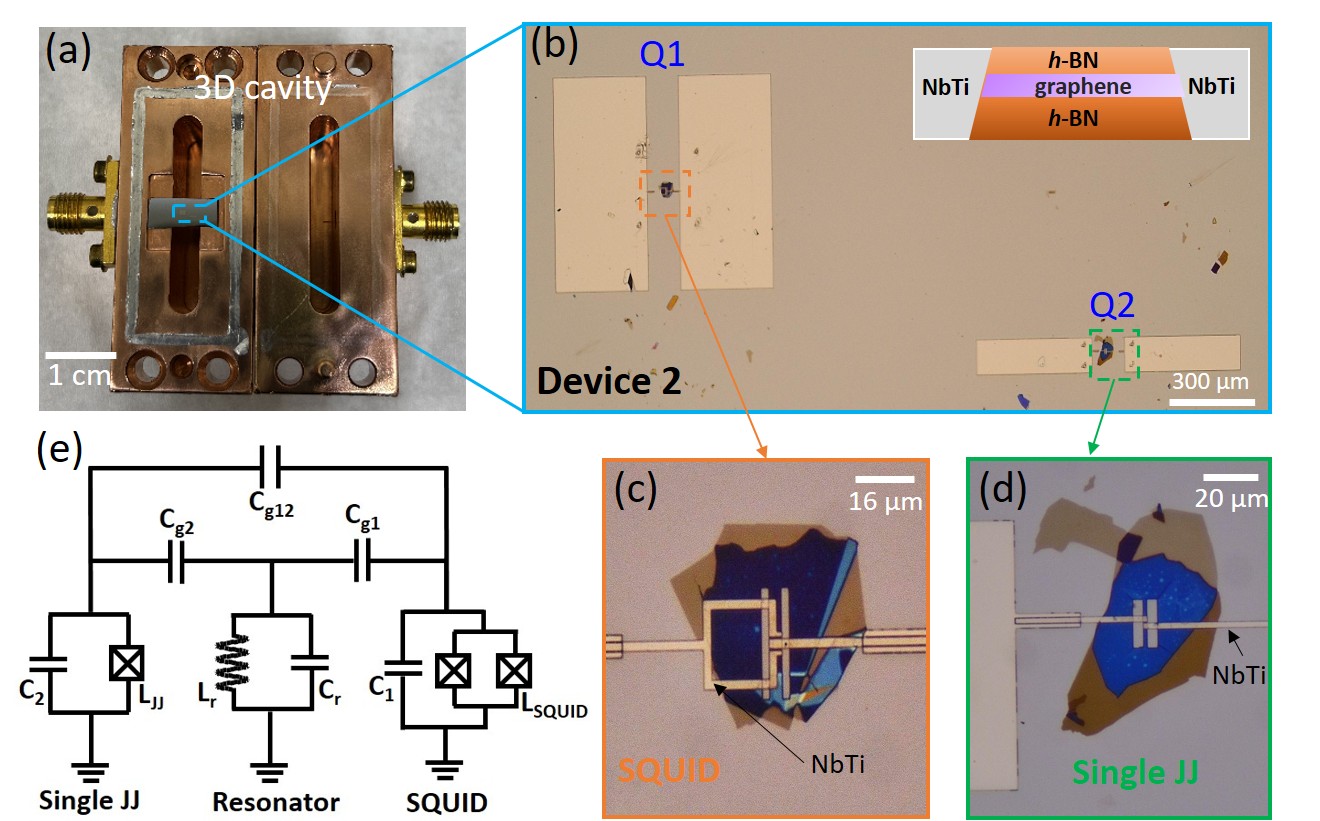}
\caption{Optical micrograph of the two-qubit graphene-based superconducting quantum circuit coupled to a 3D copper cavity. (a) Image of qubit
chip mounted in the 3D copper cavity. (b) Optical micrograph of device 2, consisting of a SQUID-based qubit (Q1) and a single-JJ qubit (Q2). The inset shows the hBN/graphene/hBN sandwich structure with NbTi edge contacts for the junctions shown in (c) and (d). (c) Optical micrograph of the SQUID circuit consisting of graphene and superconductor NbTi. The design parameters can be found in Figure \ref{FigS1}(g). (d) Optical micrograph of the single-JJ circuit. The design parameters can be found in Figure \ref{FigS1}(g). (e) Schematic of the two qubits coupled to a microwave resonator. Q1 is coupled to the resonator $via$ the capacitance $C_{g1}$, while Q2 is coupled $via$ $C_{g2}$. Q1 is coupled to Q2 $via$ $C_{g12}$.} \label{Fig1}
\end{figure}

\section{II. Device fabrication and measurement setup}
Fig. \ref{Fig1} shows the optical micrograph of one of the two devices investigated in this study. Our 3D cavity, fabricated from copper as shown in Fig. \ref{Fig1}(a), provides good thermal conductivity while allowing magnetic flux to penetrate. The graphene-based superconducting quantum circuits investigated in this work are all designed in a two-qubit layout. We present data from a two-SQUID device (device 1; see Fig. \ref{FigS1}) in which one SQUID was nonfunctional and exhibited no cavity response, as well as from a device comprising one SQUID and one single JJ (device 2). Figure \ref{Fig1}(b) shows a micrograph of device 2, where the SQUID-based qubit (Q1) is equipped with a pair of bigger capacitor pads (590 $\mu$m $\times$ 320 $\mu$m) while the single-JJ qubit (fixed qubit, Q2) is equipped with smaller capacitor pads (96.76 $\mu$m $\times$ 400 $\mu$m). Figures \ref{Fig1}(c) and \ref{Fig1}(d) show micrographs of Q1 and Q2, respectively, while Fig. \ref{Fig1}(e) presents an effective schematic illustrating how the two qubits are coupled to one resonator. Both the SQUID and the single JJ are realized using NbTi-graphene-NbTi junctions fabricated with edge-contact techniques \cite{Wang2013}. The design parameters and details of the device fabrication process are provided in section I of the Supplementary Materials (SM).

The measurement schemes employed in this work are illustrated in Fig. \ref{FigS2}. The devices were mounted in a two-port 3D copper cavity, with Q1 located at the center of the chip and aligned with the center of the cavity chamber. The SMA port located near the center of the cavity, where the chip is positioned, is used for readout, while the off-center SMA port serves as the drive port. Additional details on the 3D cavity calibration and measurement setups can be found in Ref. \cite{Chiu2025} and in section II of the SM.

\section{III. Device characterization and results}
We first characterize the SQUID-based qubit (Q1) in device 1, as shown in Figs. \ref{FigS1}(b) and \ref{FigS1}(e). This device was measured in different cavities across multiple cooldowns, as summarized in section III of the SM. Here, we begin with the data obtained from cooldown 2. Figure \ref{Fig2}(b) presents the measured S$_{21}$ through the cavity readout port as a function of readout frequency and power at zero magnetic flux [labeled (b) in Fig. \ref{Fig2} (a)]. Power-dependent cavity response provides a useful check for the presence of a qubit, as the cavity resonance shifts toward the qubit frequency at sufficiently high readout powers \cite{Krantz2020}. At low power (-50 dBm), the cavity exhibits a dip at $f_r = 6.02937$ GHz. When the power exceeds -30 dBm, the qubit becomes saturated, and the cavity shifts to its bare resonance $f_{bare} = 6.0558$ GHz, accompanied by a narrower linewidth. The readout power required for the cavity to enter this high-power regime is characterized by the critical photon number $\overline{n}_{crit}$ $\approx$ $\frac{\Delta^2}{4g^2}$, where $\Delta = 2\pi(f_{bare} - f_q)$ is the detuning between the bare cavity frequency $f_{bare}$ and the qubit frequency $f_q$, and $g$ is the qubit-cavity coupling strength \cite{Blais2004a,Gambetta2006,Boissonneault2008}. Fig. \ref{Fig2}(c) shows the power-dependent cavity response measured at another flux bias, as labeled (c) in Fig. \ref{Fig2} (a), where $f_q$ is minimized to approximately zero. As expected, with increased detuning $\Delta$ compared to the case in Fig. \ref{Fig2} (b), the onset power for the high-power regime also increases to approximately -25 dBm. Next, we perform flux tuning of the cavity response by measuring S$_{21}$ at a low readout power (-50 dBm) while sweeping the DC current applied to the superconducting coil. As shown in Fig. \ref{Fig2}(a), the flux-modulated cavity response bends both upward and downward relative to the bare cavity frequency, indicating that $f_q$ crosses $f_{bare}$ due to the tunability provided by the functional SQUID. In the vicinity of the resonant limit $f_q$ $\approx$ $f_{bare}$, the qubit and resonator hybridize, resulting in an avoided crossing as shown in Fig. \ref{Fig2}(d). The split cavity peaks at $f_q$ $\approx$ $f_{bare}$ is known for vacuum Rabi splitting, indicating that the device operates in the strong coupling regime, where the coupling strength $g$ exceeds both the qubit and cavity decay rates ($\gamma$ and $\kappa$). The frequency of the hybridized qubit-cavity states can be written as $f_{\pm}$ = $\left[f_q+f_{bare}\pm\sqrt{(f_q-f_{bare})^2+4(g/2\pi)^2}\right]/2$ \cite{Larsen2015}. Figure \ref{Fig2}(e) shows the splitting $\delta$ = $f_+ - f_-$ as a function of $f_q$, from which we can extract the coupling strength $g/2\pi$ = 100.5 MHz through fitting. Note that the extracted coupling strength $g$ agrees well with simulations and remains reasonably consistent for the same device loaded into different cavities across multiple cooldowns (see sections III and V of the SM).

\begin{figure*}[!t]	
\includegraphics[scale=0.6]{Fig2.jpg}
\caption{Flux- and power-dependent cavity response (S$_{21}$) of the single-qubit device (Q1 in device 1) in cooldown 2. (a) Flux modulation of the cavity frequency for device 1. Note that the upper part and the bottom part were measured separately in order to improve the visibility of cavity response. The arrows indicate the flux bias points where the power-dependence measurements in (b) and (c) were performed. The dashed rectangle denotes the flux and frequency ranges displayed in (d). (b) Power-dependence measurement of device 1 at the zero-flux point (sweet spot) labeled (b) in (a). (c) Same as (b), but measured at the other flux point, where $f_q$ attains its minimum, as labeled (c) in (a). (d) Zoom-in measurement of the blue rectangular region indicated in (a), showing a vacuum Rabi splitting when the qubit frequency aligns with the bare cavity frequency. Note that the data have been mean-subtracted and smoothed along the flux axis for clarity. (e) Frequency splitting between the hybridized qubit-cavity states, as a function of $f_q$, as extracted from (a).}
\label{Fig2}
\end{figure*} 

\begin{figure*}[!t]	
\includegraphics[scale=0.62]{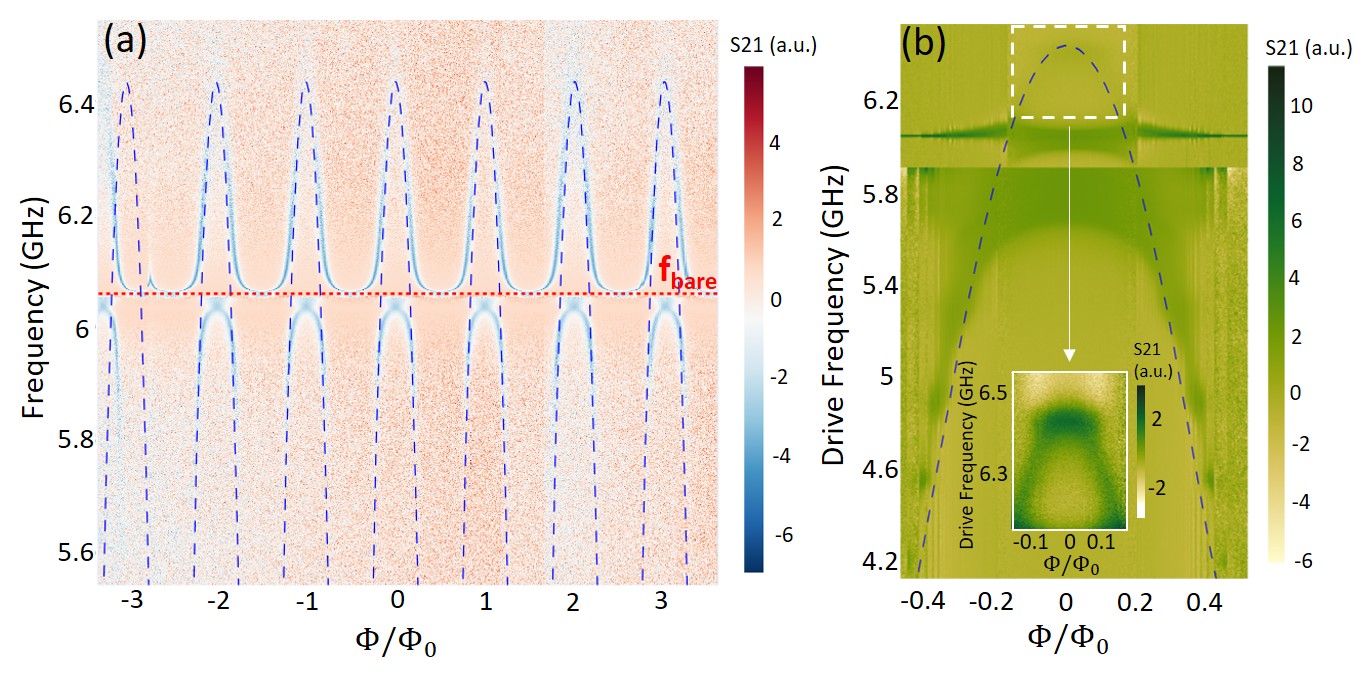}
\caption{Two-tone qubit spectroscopy of device 1 in cooldown 2. (a) Flux-dependent cavity response with the qubit frequency superimposed. Note that the data have been mean-subtracted for clarity. The dashed line indicates the inferred qubit frequency $f_q \approx f_{q,max}\sqrt{|\cos(\Phi)|}$, with a maximum value $f_{q,\mathrm{max}} \approx 6.438$ GHz as mentioned in the main text. (b) Two-tone measurement of the qubit transition as a function of applied flux $\Phi$ for device 1. The readout power was fixed at -50 dBm. The measurement was performed with two different drive powers: 15 dBm for the lower part of the spectrum, where $f_q$ is far detuned from $f_{bare}$, and -13 dBm for the upper part, where $f_q$ is close to $f_{bare}$. The spectrum is strongly distorted when the drive frequency approaches the bare cavity frequency ($f_{\mathrm{bare}} = 6.0558$ GHz). To resolve the maximal qubit frequency, a zoom-in measurement of the region indicated by the white dashed rectangle is performed and shown in the inset.}
\label{Fig3}
\end{figure*}

With the obtained $g$, we can estimate the maximum qubit frequency $f_{q,max}$ from the dispersive shift $\chi/2\pi \approx 26.4$ MHz between the low-power ($f_r$) and high-power ($f_{bare}$) cavity frequencies [Fig. \ref{Fig2}(b)], using $\chi = g^2/\Delta$ and $\Delta = 2\pi(f_{q,max} - f_{bare})$, yielding $f_{q,max} \approx 6.438$ GHz. The symmetric SQUID qubit can therefore be modeled as $f_q \approx f_{q,\mathrm{max}} \sqrt{|\cos(\Phi)|}$ \cite{Chiu2025}, which we overlay on the flux-modulated cavity response in Fig. \ref{Fig3}(a). To verify the modeled qubit spectrum, we performed two-tone spectroscopy in continuous-wave mode, where a second continuous drive tone was applied to the cavity in addition to the readout tone. The measurement was carried out by recording S$_{21}$ at the cavity’s resonant frequency for different magnetic flux values while sweeping the drive-tone frequency. When the drive tone was on resonance with qubit frequency $f_q$ at each flux point, a peak in the cavity response was observed, yielding a flux-tunable qubit transition spectrum as shown in Fig. \ref{Fig3}(b). The spectrum near the bare cavity frequency ($f_{bare}$ = 6.0558 GHz) is distorted because the drive tone, being close to the cavity resonance, strongly excites the cavity mode and saturates the readout signal. Therefore, a fine scan excluding this range and using low drive power (-13 dBm) was performed to accurately determine $f_{q,max}$, as shown in the inset of Fig. \ref{Fig3}(b). The obtained $f_{q,max}$ $\approx$ 6.43 GHz is in excellent agreement with the estimated value based on $g$ and dispersive shift. 

\begin{figure*}[!t]	
\includegraphics[scale=0.58]{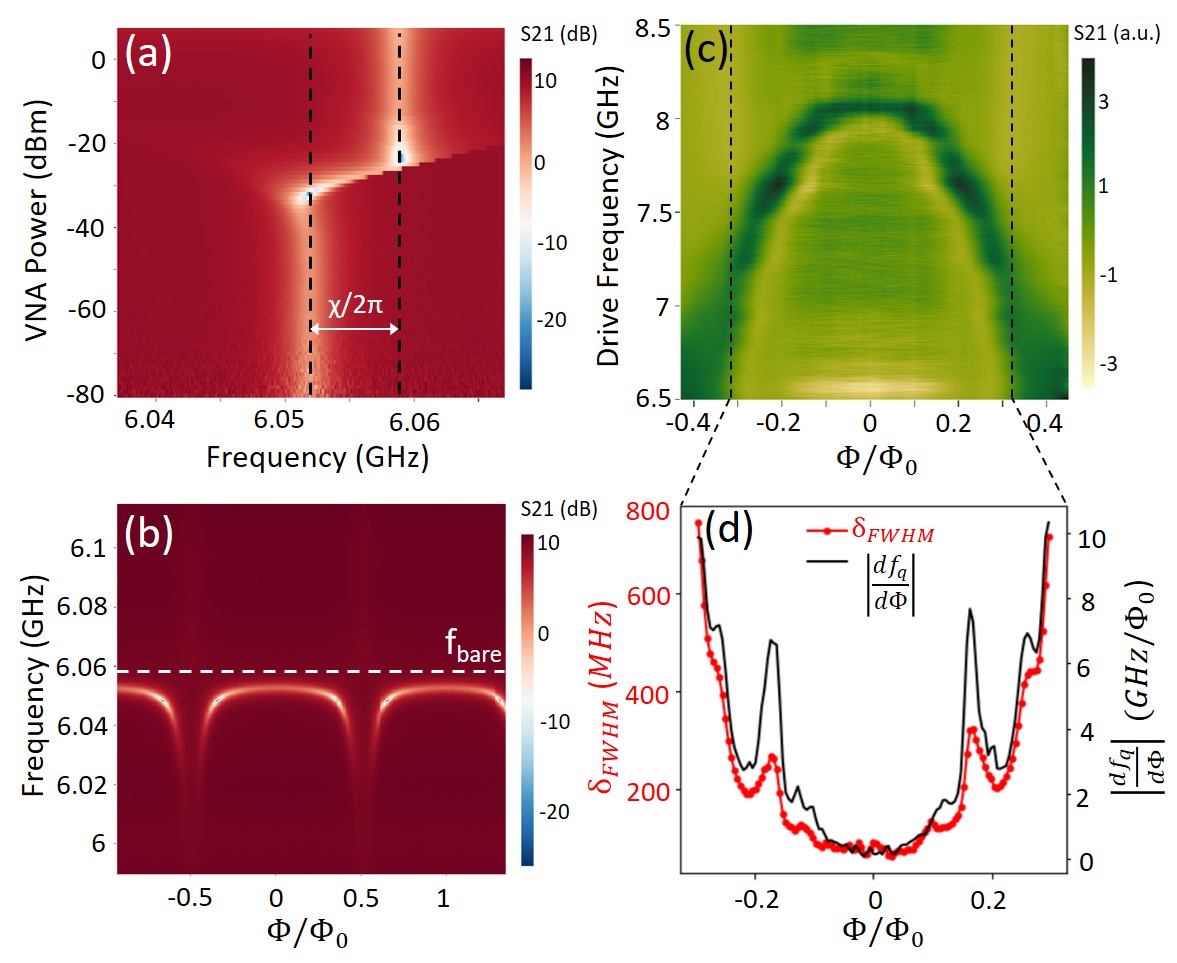}
\caption{Power dependence, flux-tuning and qubit spectroscopy for device 1 in cooldown 1. (a) Power dependence measurements of device 1 performed at $\Phi$ = 0 (sweet spot), with a dispersive shift $\chi/2\pi \approx 6.15$ MHz. (b) Flux modulation of the cavity frequency measured with a readout power of -60 dBm. The white dashed line indicates the bare cavity frequency at 6.059 GHz. (c) Two-tone measurement of the qubit transition as a function of applied flux, measured with a readout power of -60 dBm and a drive power of 8 dBm, showing a maximal qubit frequency $f_{q,\mathrm{max}} \approx 8.068$ GHz. The dashed lines denote the flux range for the analysis shown in (d). (d) Correspondence between the spectral linewidth ($\delta_{FWHM}$) and the flux derivative of qubit frequency ($\left|df_q/d\Phi\right|$) extracted from the spectrum in (c).}
\label{Fig4}
\end{figure*}

Next, we examine the flux dependence of the spectral broadening, which reflects the qubit coherence time $T_2^\ast$ \cite{Schuster2005,Bishop2009,AlbertHertel2022}. The spectrum in Fig. \ref{Fig3}(b), however, is unsuitable for this analysis because of its distortion around $f_{bare}$. In contrast, during another cooldown of the same device, shown in Fig. \ref{Fig4}, the maximum qubit frequency lies well above $f_{bare}$, enabling us to extract an undistorted spectrum. This highlights a key advantage of 3D cavity devices, where switching cavities with different resonant frequencies can be employed to probe various qubit-cavity coupling regimes for the same device. Fig. \ref{Fig4}(a) and (b) show the power dependence and flux tuning of the cavity response, while Fig. \ref{Fig4}(c) present the qubit transition frequency as a function of flux, for this cooldown. Using $\chi = g^{2}/\Delta$, the dispersive shift $\chi/2\pi \approx 6.15$ MHz at $\Phi = 0$ [Fig. \ref{Fig4}(a)] together with the maximum qubit frequency $f_{q,\mathrm{max}} \approx 8.068$ GHz [Fig. \ref{Fig4}(c)] yields a coupling strength $g/2\pi \approx 111.3$ MHz. This value is consistent with the vacuum Rabi splitting result $g/2\pi = 100.5$ MHz from cooldown 2. The half linewidth of the $\left|0\right\rangle$ to $\left|1\right\rangle$ qubit transition ($\delta_{HWHM}$) is related to the coherence time $T^\ast_2$ and the relaxation time T$_1$ $via$: $(2\pi\delta_{HWHM})^2$ = $(\frac{1}{T^\ast_2})^2+n_s\omega_{vac}^2\frac{T_1}{T^\ast_2}$, where $n_s$ is the average number inside the resonator, $\omega_{vac} = 2g$ is the vacuum Rabi frequency, and $n_s\omega_{vac}^2$ is proportional to the drive power \cite{Sun2023,Aparicio2025}. The full width at half maximum (FWHM) of the spectral peak, extracted from Fig. \ref{Fig4}(c), is shown in red in Fig. \ref{Fig4}(d). Since the spectrum in Fig.~\ref{Fig4}(c) was measured under a finite drive power of 8 dBm (nonzero $n_s \omega_{\mathrm{vac}}^2$), the observed spectral linewidths, which range from approximately 64.6 MHz at $\Phi = 0$ to 743.7 MHz at $\Phi = \pm 0.32 \Phi_0$, provide a lower-bound range for $T_2^\ast$ from 4.927 ns to 0.428 ns, using $T_2^\ast \approx 1/(2\pi\delta_{\mathrm{HWHM}})$. We also performed a power-broadening measurement at $\Phi = 0$ (the sweet spot) and extracted an estimated $T_2^\ast \approx$ 17.63 ns, as discussed in section IV of the SM. The linewidth in Fig. \ref{Fig4}(c) is narrowest at the sweet spot, where $df_q/d\Phi$ approaches zero, and becomes progressively broader away from the sweet spot as $df_q/d\Phi$ increases. In Fig. \ref{Fig4}(d), we plot $\left|df_q/d\Phi\right|$ extracted from Fig. \ref{Fig4}(c), and find a strong correlation between $\delta_{FWHM}$ and $df_q/d\Phi$. Since $df_q/d\Phi$ determines the sensitivity of the qubit frequency to low-frequency flux noise, this correlation suggests that $T_2^\ast$ is dominated by qubit frequency fluctuations caused by low-frequency noise \cite{Ithier2005} in this cooldown. By using $\delta f_{\mathrm{FWHM}} = \left| \frac{d f_q}{d \Phi} \right| A \sqrt{\ln 2}$ and restricting the analysis to the vicinity of the sweet spot ($|\Phi/\Phi_0| \lesssim 0.05$) \cite{Nguyen2019}, we extract an upper bound on the flux-noise amplitude $A \sim 10^{-1}\,\Phi_0$, which is four orders of magnitude larger than typical values reported for conventional superconducting qubits \cite{Leon2021}. In addition, we performed time-domain pulse measurements to probe the qubit relaxation time, yielding $T_1 \approx 48$ ns (see section IV of the SM), consistent with values reported for graphene gatemons \cite{Wang2019}.

\begin{figure*}[!t]	
\includegraphics[scale=0.52]{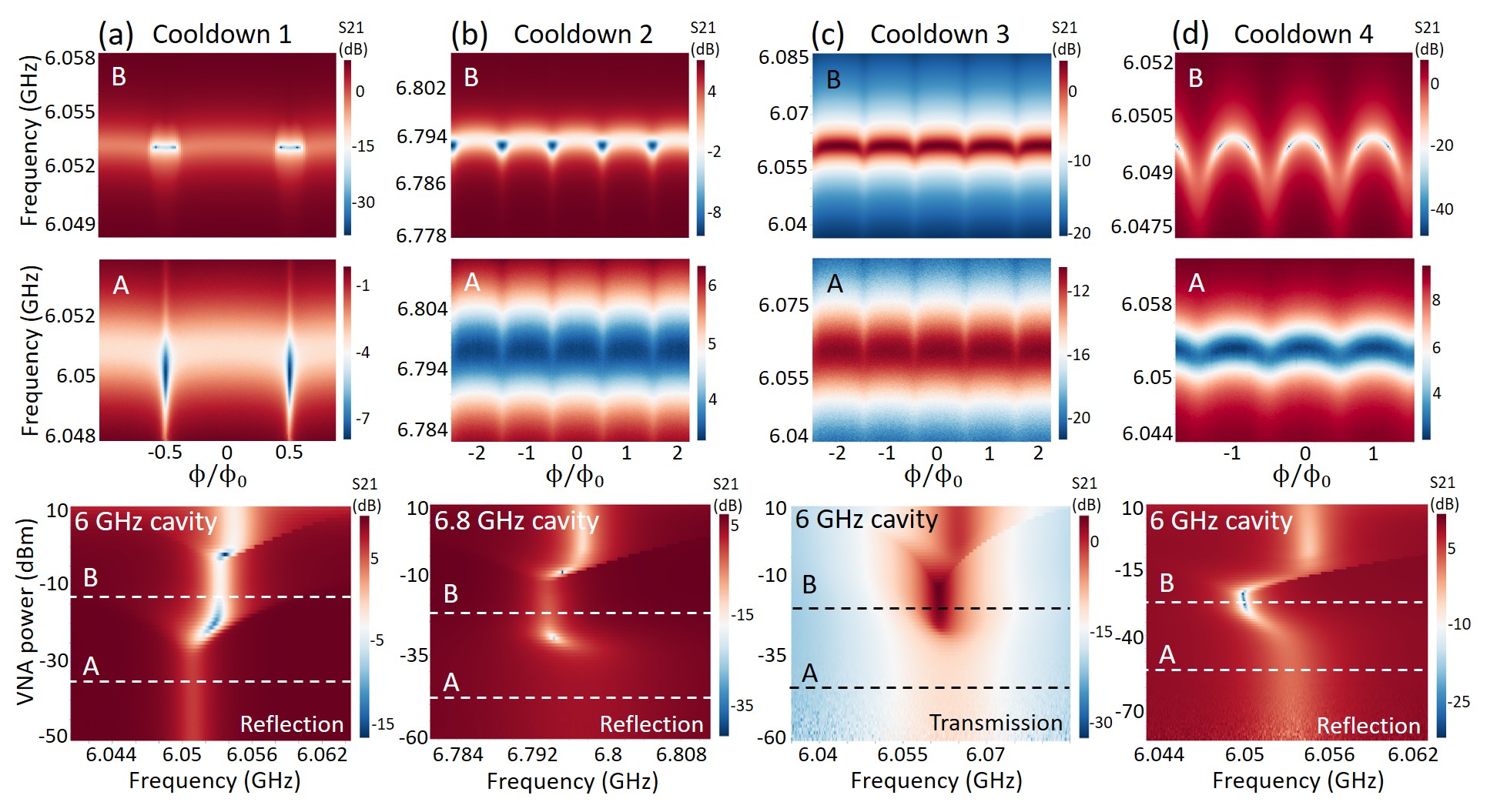}
\caption{Device 2 loaded into different cavities over multiple cooldowns. The first, second, and fourth cooldowns were performed using reflection measurements, as shown in Fig. \ref{FigS2}, whereas the third cooldown employed transmission measurements, as described in Ref. \cite{Chiu2025}. (a) Power dependence of device 2 in a 6 GHz cavity, measured at the sweet spot (bottom panel), together with the flux modulation of the cavity response measured using VNA powers labeled A (middle panel) and B (top panel) in the bottom panel. (b) Same as (a), but measured in a 6.8 GHz cavity. (c) Same as (a), but using a transmission measurement setup. (d) Same as (a).}      
\label{Fig5}
\end{figure*}

Having addressed the single-qubit device, we now turn to the two-qubit device (device 2), consisting of one SQUID-based circuit and one single-JJ circuit, as shown in Fig. \ref{Fig1} (a) - (d). We loaded device 2 in 6 GHz and 6.8 GHz cavities and performed reflection or transmission measurements across four cooldowns, as shown in  Fig. \ref{Fig5}. In Fig. \ref{Fig5}, the bottom panels show the power-dependence measurements performed at the sweet-spot flux point, while the top and middle panels present the flux-modulated cavity response recorded at the two readout powers, labeled B and A in the bottom panels, across different cooldowns. A two-stage dispersive shift is clearly observed in all power-dependence measurements, indicating the presence of two qubits coupled to a single cavity mode, with each qubit saturating at different power levels. The measured flux modulation of the cavity response at two different readout powers reveal the origin of each dispersive shift. As can be seen in all the middle panels, the measured spectrum exhibits the flux-dependent oscillations characteristic of a SQUID-based qubit. After passing the first dispersive-shift threshold, the measured spectrum shows that the oscillatory behavior persists, as seen in the top panel of Fig. \ref{Fig5}. These observations indicate that the first dispersive shift originates from the single-JJ circuit (i.e., the fixed qubit), as the flux-dependent modulation remains visible beyond the first dispersive shift. Additional evidence supporting this assignment is that the onset power for the first dispersive shift is fixed, whereas that for the second dispersive shift varies with flux, as shown in Fig. \ref{FigS5} (see discussions in section VI of SM). Furthermore, the fact that the fixed qubit saturates at a lower power, while the SQUID-based qubit does so at a higher power, is also discussed in section VI of SM. 

The first dispersive shift in the bottom panel of Fig. \ref{Fig5} (a) indicate $f_q$ of the fixed qubit is larger than 6 GHz in the first cooldown, while that in the bottom panel of Fig. \ref{Fig5} (b)-(d) indicate $f_q$ varies below 6.8 GHz and 6 GHz in the rest of the cooldowns. On the other hand, the qubit frequency for SQUID-based qubit is always higher than $f_{bare}$, as the second dispersive shifts are all toward higher frequency in all cooldowns [see the bottom panels of Fig. \ref{Fig5}]. Since the capacitor pad design of the SQUID circuit is identical in devices 1 and 2, we use the experimentally obtained coupling strength of $g/2\pi \approx 100$ MHz from device 1 to estimate $f_{q,\mathrm{max}} = 13.86$ GHz for device 2, based on the second dispersive shift of 1.28 MHz extracted from Fig. \ref{FigS5}(b) in cooldown 1. This inferred $f_{q,\mathrm{max}}$ is consistent with our junction design, where the junction width in device 2 (3.5~$\mu$m) is larger than that in device 1 (1~$\mu$m) [see Fig. \ref{FigS1}(g)]. Given that $f_q \approx \sqrt{8E_JE_C}/h$ and $E_J = \Phi_0 I_C / 2\pi$, we expect $f_{q,\mathrm{max}}$ for device 2 to scale as $\sqrt{3.5}$ times that of device 1, yielding $\sqrt{3.5} \times$ 8.068 GHz $\approx$ 15.1 GHz. In section V of SM, we simulated a coupling strength of $g/2\pi \approx 78.9$ MHz for the fixed qubit. Using this value, we further estimate the qubit frequency of the fixed qubit from the first dispersive shift observed in each cooldown. The corresponding discussions are provided in section VI of SM. It is worth noting that the linewidth of the cavity response after the first dispersive shift is reduced, as seen in the bottom panels, or by comparing the flux modulation data in the middle and upper panels of Fig. \ref{Fig5}. This suggests that the fixed qubit acts as a loss channel for the cavity. Once the fixed qubit is saturated, the cavity effectively eliminates this lossy channel and regains a narrower linewidth. We have performed continuous-wave two-tone measurements on device 2. No flux-tunable qubit transition was detected, as was achieved in device 1, but potential fixed-qubit-related transitions are present, as discussed in section VI of the SM. The fact that $f_q$ of the SQUID-based qubit crosses the cavity periodically but does not form avoided crossings, indicates the device is not in the strong coupling regime (i.e., $\Omega$ $\gg$ $\gamma$, $\kappa$, where $\Omega$ is the qubit-cavity energy exchange rate and $\gamma$($\kappa$) is qubit (cavity) decay rate). This suggests a short coherence time for the SQUID-based qubit in device 2, with decay rate $\gamma$ $>$ $\Omega$ = $g/\pi$ $\approx$ 200 MHz, corresponding to a coherence time 1/$\gamma$ $<$ 5 ns, which may prevent us from resolving the qubit transition peaks in two-tone spectroscopy measurements. Comparing device 1 and device 2, the coherence properties of our qubits still exhibit device-to-device variations, likely arising from variations in the quality of the constituent graphene junctions. Nevertheless, with the two distinct dispersive shifts demonstrated, together with future improvements in coherence properties, it will be possible to read out two qubits and study qubit-qubit interactions in our system.

\maketitle
\section{IV. Summary}
In this work, we have constructed single- and two-qubit graphene-based superconducting quantum circuits coupled to 3D cavities. For both devices, we placed them in cavities with different resonant frequencies, allowing us to access distinct coupling regimes between the qubits and the resonators. We present the first flux-tunable graphene transmon, with a measured $T_1$ $\approx$ 48 ns and a lower bound estimate of $T_2^\ast$ $\approx$ 17.63 ns. The single-qubit device has demonstrated a strong coupling regime when the qubit frequency is tunable across the bare cavity frequency. In another cooldown, where the qubit frequency lies well above the bare cavity resonance, we observe a flux-dependent spectral linewidth, suggesting that $T_2^\ast$ is limited by qubit frequency fluctuations induced by low-frequency flux noise. In the two-qubit device, we observed a distinct two-stage dispersive shift in the power-dependent measurements across four cooldowns, indicating variations in the detuning among the fixed qubit, the SQUID-based qubit, and the resonator when the device was placed in different cavities. The first dispersive shift is attributed to the fixed qubit, while the second corresponds to the SQUID-based qubit, as supported by the flux-tunable cavity response observed after the first dispersive shift and the flux-dependent onset power of the second. In summary, we demonstrate the flexibility in qubit-cavity coupling achievable by 2D-material-based transmons when integrated with 3D cavities. The two-qubit architectures highlight the prospects for constructing multiqubit 3D transmon based on 2D materials for joint readout and control \cite{Filipp2009}.

\maketitle
\section{V. Acknowledgments}
Kuei-Lin Chiu would like to thank the funding support from National Science and Technology Council (Grant No. NSTC 113-2112-M-110-015). Chung-Ting Ke would like to thank the funding support from National Science and Technology Council (Grant No. NSTC 110-2628-M-001-007). Kuei-Lin Chiu and Yen-Hsiang Lin also acknowledge support from the Center for Quantum Science and Technology (CQST) within the framework of the Higher Education Sprout Project by the Ministry of Education (MOE) in Taiwan.

\maketitle
\section{VI. Author contributions}
K. L. Chiu conceived the project. Y. H. Lin designed the qubit layouts. C. H. Lo, A. J. Lasrado, Y. C. Chen, and S. P. Shih fabricated the devices under the supervision of K. L. Chiu and C. T. Ke. A. J. Lasrado, C. H. Lo, and Y. C. Chen performed the measurements under the supervision of K. L. Chiu. A. J. Lasrado performed the HFSS simulations. K. L. Chiu wrote the manuscript with input from all co-authors. 

\maketitle
\section{VII. Competing financial interests}
The authors declare no competing financial interests.

 \widetext
 \clearpage
 \begin{center}
 	\textbf{\large Supplementary materials}
 \end{center}

 \setcounter{section}{0}
 \setcounter{equation}{0}
 \setcounter{figure}{0}
 \setcounter{table}{0}
 \setcounter{page}{1}
 \makeatletter
 \renewcommand{\thesection}{S\arabic{section}}
 \renewcommand{\thesection}{\Roman{section}}
 \renewcommand{\theequation}{S\arabic{equation}}
 \renewcommand{\thefigure}{S\arabic{figure}}
 \renewcommand{\bibnumfmt}[1]{[S#1]}

 \section{I. Device fabrication}

\begin{figure}[h]
 	\includegraphics[scale=0.65]{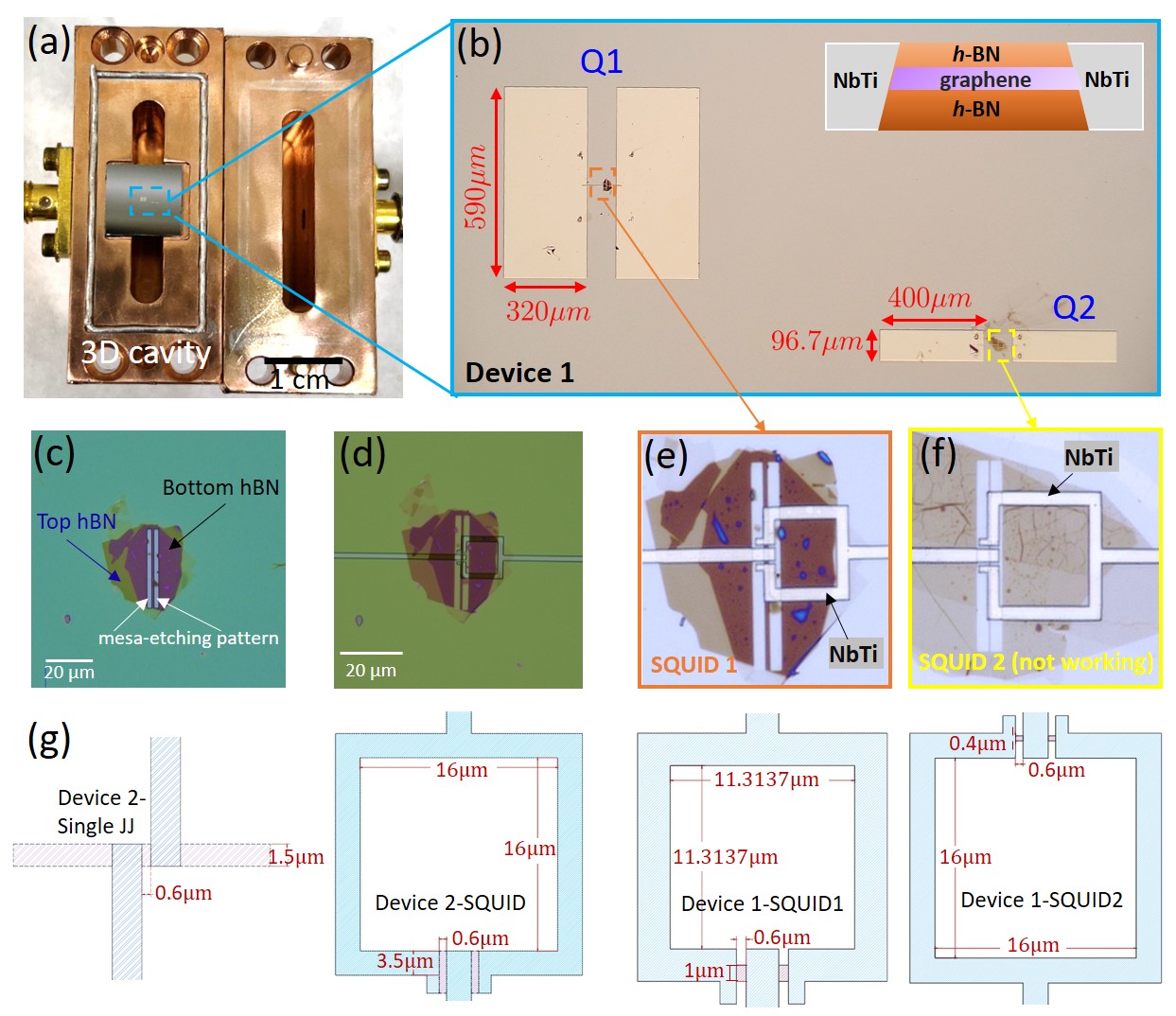}
 	\caption{Optical microscope images of device 1 at different stages during fabrication. (a) Image of qubit chip (device 1) mounted in the 3D copper cavity. (b) Optical micrograph of device 1, consisting of two SQUID-based qubits labeled Q1 and Q2, where Q2 is non-functional. (c) Optical micrograph of hBN/graphene/hBN sandwich structure of Q1 with the mesa etching pattern to define the width of graphene JJ. (d)  Optical micrograph of device 1 after mesa etching and before NbTi sputtering. (e) Optical microscope image of the as-fabricated Q1 (SQUID1) in device 1. (f) Optical microscope image of the as-fabricated Q2 (SQUID2), which is not functional, in device 1. (g) The design parameters for Q1 and Q2 in device 1 and device 2. The red regions denote the graphene weak links, while the light blue regions represent their superconducting contacts.}      
 \label{FigS1}
 \end{figure}

Figures \ref{FigS1}(a)–(f) show optical micrographs of the as-fabricated device 1 and the constituent qubit Q1 at different stages of fabrication. Here, we take Q1 in device 1 as an example to elaborate the fabrication process. We first prepare the hBN/graphene/hBN sandwich on intrinsic Si wafers capped with 90 nm SiO$_2$ (substrate) using the polymer-free dry transfer method \cite{R.2010,Mayorov2011,Haigh2012,Wang2013}. After spin coating of PMMA (A6, 500 rpm for 5 s then 4000 rpm for 55 s, bake at 170 $^\circ$C for 2 mins), electron-beam lithography (EBL) was used to define the mesa-etching pattern that determines the junction width. After EBL exposure and developing, a Inductively Coupled Plasma Reactive-Ion Etching (ICP-RIE) technique was performed, using CHF$_3$ and O$_2$ gases with a ratio of 20:1 (power: 150 W and bias: 20 V), to selectively etch the stack and define the junction width, as shown in Fig. \ref{FigS1} (c). Subsequently, another EBL step was used to define the patterns for the SQUID contacts and capacitor pads, as shown in Fig. \ref{FigS1}(d). To establish edge contact with the encapsulated graphene, another ICP-RIE process was carried out, followed by sputtering of a 120 nm NbTi layer (pressure: 3 mTorr, power: 25 W, sputtering rate: 10 nm/min). The as-fabricated Q1 and Q2 in device 1 are shown in Fig. \ref{FigS1}(e) and Fig. \ref{FigS1}(f), respectively. The capacitor pads in Q1 measure 590 $\mu$m $\times$ 320 $\mu$m, whereas those in Q2 are 400 $\mu$m $\times$ 96.7 $\mu$m in both devices 1 and 2. Finally, the design parameters for Q1 and Q2 for both devices are shown in Fig. \ref{FigS1}(g).

\section{II. Measurement scheme}

All the experiments detailed in the main text were performed in a dilution refrigerator with a base temperature of 10 mK. In this section, we introduce our 3D cavities and describe the measurement setups used in this work.

\begin{figure}[!t]	
 	\includegraphics[scale=0.65]{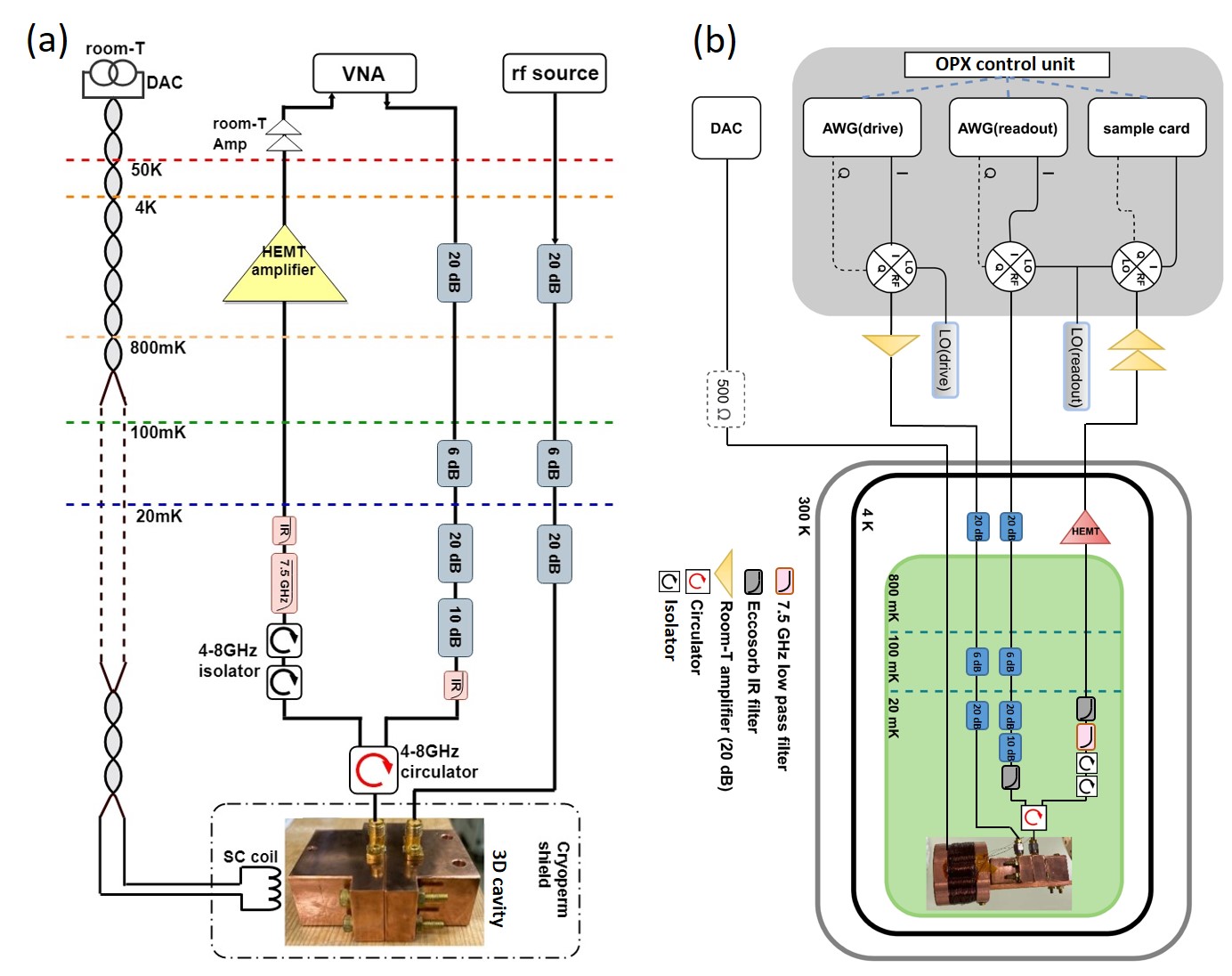}
 	\caption{(a) Measurement setup used for frequency-domain measurements, including power dependence, flux modulation, and two-tone spectroscopy of the graphene-based superconducting quantum circuits. (b) Measurement setup used for time-domain pulse measurements. The main control and readout system is an OPX-1000 module manufactured by Quantum Machines.}      
 	\label{FigS2}
 \end{figure}

\subsection{3D cavity setup}

The design and calibration of the 3D cavities can be found in our previous work \cite{Chiu2025}. The transmon chips positioned at the center of the cavity chamber [Fig. \ref{Fig1} (a)] are primarily couple to the electric field of the TE$_{101}$ mode. The 3D copper cavity is milled from two pieces, one half consisting of a pin to drive qubits (drive port), and another with a pin to read qubits (readout port). The readout port is located closer to the center of the cavity (where the chip is mounted) than the drive port, as shown in the bottom of Fig. \ref{FigS2}. The readout port is intentionally overcoupled using a longer pin to ensure strong interaction with the cavity field, whereas the drive port employs a shorter pin, resulting in weaker coupling, a higher external quality factor ($Q_{ext}$), and reduced photon leakage. For empty cavities, the internal quality factor ($Q_{int}$) typically ranges between 3000 and 3500 at room temperature and increases to approximately 20,000 at 10 mK. For devices 1 and 2, the $Q_{ext}$ of the readout port was calibrated to 1000 in all cooldowns. The $Q_{\mathrm{ext}}$ of the drive port in both devices was calibrated to values between 6000 and 7000 for cooldowns 2, 3, and 4. In contrast, during cooldown 1 for both devices, the drive and readout operations were performed through the same readout pin ($Q_{\mathrm{ext}} \approx 1000$). 

\subsection{Measurement setups}
The measurements were performed in a dilution refrigerator, where the 3D copper cavities were thermally anchored to the mixing chamber plate ($\approx$ 10 mK). Fig. \ref{FigS2} (a) shows the reflection setup for frequency-domain measurements. The readout microwave signal from the output of a vector network analyzer (VNA, Keysight E5071C) passes through a series of low-temperature attenuators (total attenuation $\approx$ 56 dB) and an Eccosorb IR filter before reaching the first port of a circulator. The second port of the circulator is connected to cavity’s readout port. The reflected readout signal exits through the third port of circulator and enters the readout-out line, which includes two isolators in series, a low-pass filter (cutoff frequency 7.5 GHz), and an Eccosorb IR filter, before reaching a high-electron-mobility transistor (HEMT, LNF-LNC4 8C) amplifier at the 4 K stage. The HEMT provides 40 dB of gain with a noise temperature of approximately 2 K. The signal is further amplified by two room-temperature amplifiers (20 dB gain each) before going into the input of the VNA. For qubit excitation, an RF source (Rohde $\&$ Schwarz SGS100A) delivers the drive signal through a separate drive line with a total attenuation of 46 dB, connected to the cavity’s drive port. Magnetic flux biasing is provided by a home-built superconducting coil powered by an external DC source (digital-to-analog converter, DAC).

For time-domain measurements, we employ a standard heterodyne detection scheme, as illustrated in Fig. \ref{FigS2} (b). At room temperature, the readout circuit consists of an IQ mixer that combines a 4–8 GHz local oscillator (LO) signal from an RF source with a waveform generated by an arbitrary waveform generator (AWG). The resulting up-converted microwave pulse is sent through the readout-in line to probe the qubit. The reflected signal from the qubit is collected from the readout-out line, down-converted, and digitized by a signal acquisition card. For qubit control, another IQ mixer combines the drive LO with a Gaussian-shaped waveform from the AWG. The resulting signal is sent through the drive line (46 dB attenuation) to the drive port of 3D cavity. All pulse sequences are synchronized using a trigger signal derived from a rubidium frequency standard.

\section{III. Different detuning between qubit and cavity for device 1}

\begin{figure}[!t]	
\includegraphics[scale=0.5]{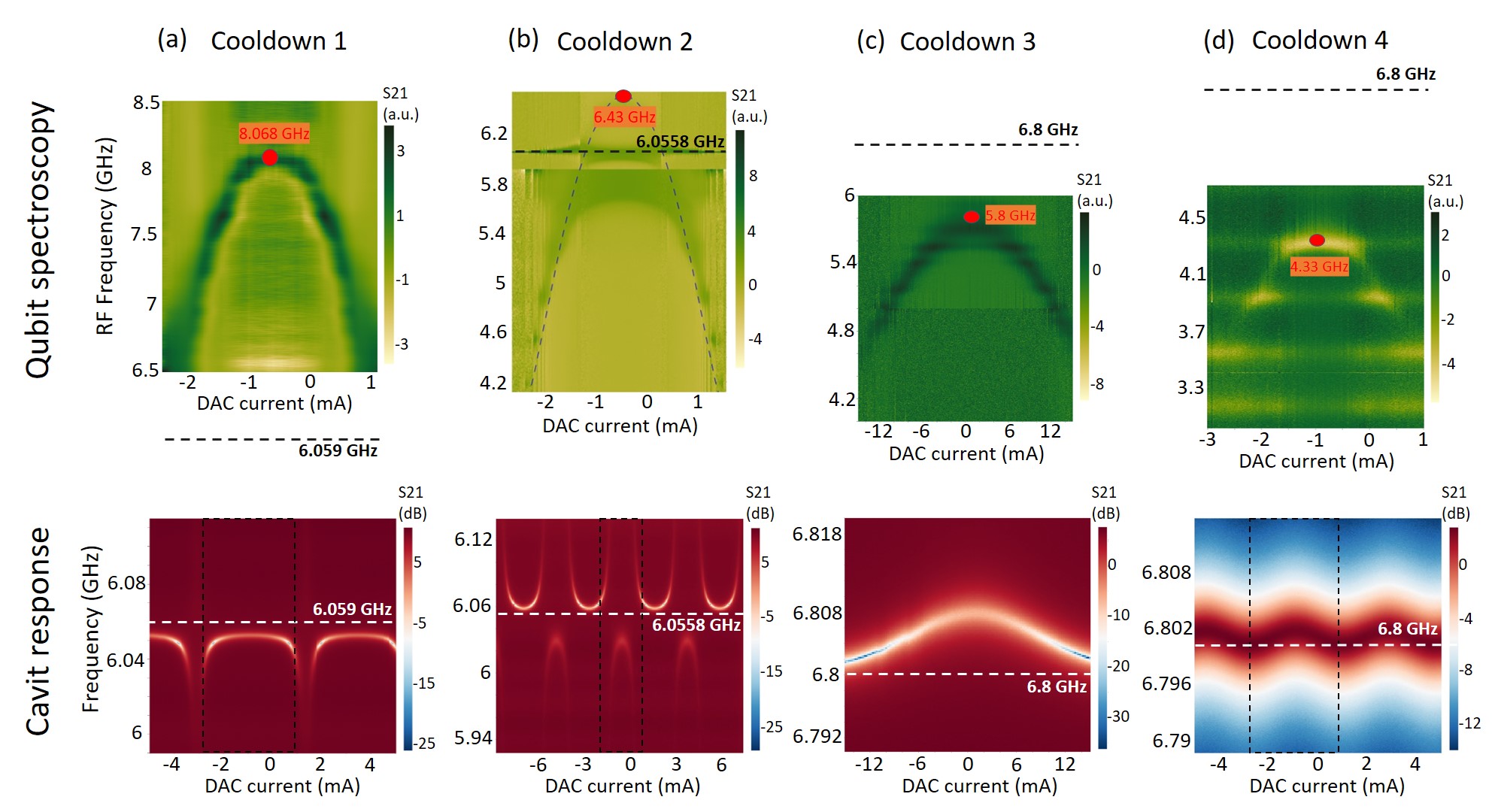}
\caption{Device 1 loaded into different cavities over multiple cooldowns. The first three cooldowns were performed using reflection measurements, as shown in Fig. \ref{FigS2}, while the fourth cooldown employed transmission measurements, as described in Ref. \cite{Chiu2025}. (a) Device 1 loaded into a 6.059 GHz cavity in the first cooldown. The bottom panel shows the cavity response as a function of flux, while the top panel presents the two-tone qubit spectroscopy measured within the rectangular region indicated in the bottom panel. The dashed lines in both panels denote the bare cavity frequency. The extracted $f_{q,\mathrm{max}} \approx 8.068$ GHz corresponds to a regime where the qubit frequency lies well above the cavity mode. (b) Same as (a) but in the second cooldown. The extracted $f_{q,\mathrm{max}} \approx 6.43$ GHz corresponds to a regime in which the qubit frequency lies just above the cavity mode. (c) Device 1 loaded into a 6.8 GHz cavity in the third cooldown. The extracted $f_{q,\mathrm{max}} \approx 5.8$ GHz corresponds to a regime where the qubit frequency lies just below the cavity mode. (d) Same as (c) but using a transmission measurement setup in the fourth cooldown. The extracted $f_{q,\mathrm{max}} \approx 4.33$ GHz corresponds to a regime where the qubit frequency lies far below the cavity mode. Note that the qubit spectra in the upper panels of (c) and (d) display oscillatory features along the frequency axis, which we attribute to standing waves in the drive lines. Consequently, we do not analyze the flux-dependent spectral linewidth for these spectra as was done in Fig. \ref{Fig4}(d).}      
\label{FigS3}
\end{figure}

Device 1 was loaded into a 6 GHz cavity for measurements during its first and second cooldowns, and into a 6.8 GHz cavity for its third and fourth cooldowns. For cooldowns 1-3, a reflection measurement setup was used, as shown in Fig. \ref{FigS2}, whereas a transmission measurement setup was employed in the fourth cooldown, as described in Ref. \cite{Chiu2025}. Figure \ref{FigS3} shows the qubit frequency and cavity response as functions of magnetic flux when device 1 was mounted in different cavities during separate cooldowns. From the qubit spectroscopy, we observe a monotonic decrease in $f_{q,\mathrm{max}}$ across cooldowns, from 8.068 GHz to 6.43 GHz, 5.8 GHz, and 4.33 GHz. Together with the corresponding bare cavity frequencies, four distinct detuning regimes between $f_{q,\mathrm{max}}$ and $f_{\mathrm{bare}}$ were achieved. In cooldown 1, the cavity response exhibit an down-bending behavior, as shown in Fig. \ref{Fig4} (b), indicating the qubit frequency approaches the bare cavity frequency from above. The measured qubit spectrum, shown in Fig. \ref{Fig4}(c), confirms that $f_{q,\mathrm{max}} \approx 8.068$ GHz is well above $f_{\mathrm{bare}} \approx 6.059$ GHz, satisfying the conditions of the strong dispersive regime, where $g/\Delta \ll 1$ and $\left| g^2/\Delta \right| \gg \kappa$. Here, $\Delta/2\pi$ $=$ $f_q - f_{bare}$, $g/2\pi = 111.3$ MHz, and $\kappa/2\pi = 2.6$ MHz is the cavity decay rate extracted from the linewidth of the low-power cavity response in Fig. \ref{Fig4} (a). This detuning regime allows us to probe the qubit spectrum around $f_{q,\mathrm{max}}$, as shown in Fig. \ref{Fig4}(c), without concern that the drive field gets too close to $f_{\mathrm{bare}}$ and distorts the spectrum. In cooldown 2, $f_{q,\mathrm{max}} \approx 6.4$ GHz lies just above $f_{\mathrm{bare}} \approx 6.0558$ GHz, causing $f_q$ to intersect with $f_{\mathrm{bare}}$ within the probed flux range. At the intersection points, the strong-coupling condition is satisfied - i.e., $\Delta \approx 0$ and $g \gg \gamma, \kappa$, where $\gamma$ is the qubit decay rate - resulting in the vacuum Rabi splitting observed in Fig. \ref{Fig2}(d). However, away from $f_{bare}$, the strong dispersive condition can still be satisfied, resulting in the resolvable qubit spectrum near $f_{q,\mathrm{max}} \approx 6.43$ GHz, as shown in the inset in Fig. \ref{Fig3}(b). In cooldown 3 and 4, cavity responses exhibit an up-bending behavior, as shown in the bottom panel of Fig. \ref{FigS3} (c) and (d), indicating the qubit frequency is approaching the bare cavity frequency from below. The measured qubit spectra, shown in the upper panels in Fig. \ref{FigS3} (c) and (d), with $f_{q,\mathrm{max}} \approx 5.8$ GHz and 4.33 GHz both below $f_{\mathrm{bare}} \approx 6.8$ GHz, confirm this speculation. In total, we have achieved four distinct regimes in this work: $f_{q,\mathrm{max}}$ far above $f_{\mathrm{bare}}$ (cooldown 1), just above $f_{\mathrm{bare}}$ (cooldown 2), just below $f_{\mathrm{bare}}$ (cooldown 3), and far below $f_{\mathrm{bare}}$ (cooldown 4). We also extract the coupling strength $g$ in each cooldown by combining the dispersive shift obtained from the power-dependence measurements at the sweet spot with the $f_{q,\mathrm{max}}$ values extracted from the qubit spectra. The measured dispersive shifts are 6.15 MHz, 26.4 MHz, 9.92 MHz, and 3.1 MHz for cooldowns 1, 2, 3, and 4, respectively, yielding $g/2\pi \approx 111.3$ MHz, 100.5 MHz, 99.98 MHz, and 87.54 MHz. The extracted coupling strength $g$ is consistent across cooldowns and agrees reasonably with our previous simulated value of 98 MHz for a qubit with the same design \cite{Chiu2025}. Note that the dispersive shifts for cooldowns 1 and 2 are shown in Fig. \ref{Fig4}(a) and Fig. \ref{Fig2}(b), while the data for cooldowns 3 and 4 are not shown here.

\section{IV. Coherence time measurements and analysis for device 1}

\begin{figure}[!t]	
\includegraphics[scale=0.55]{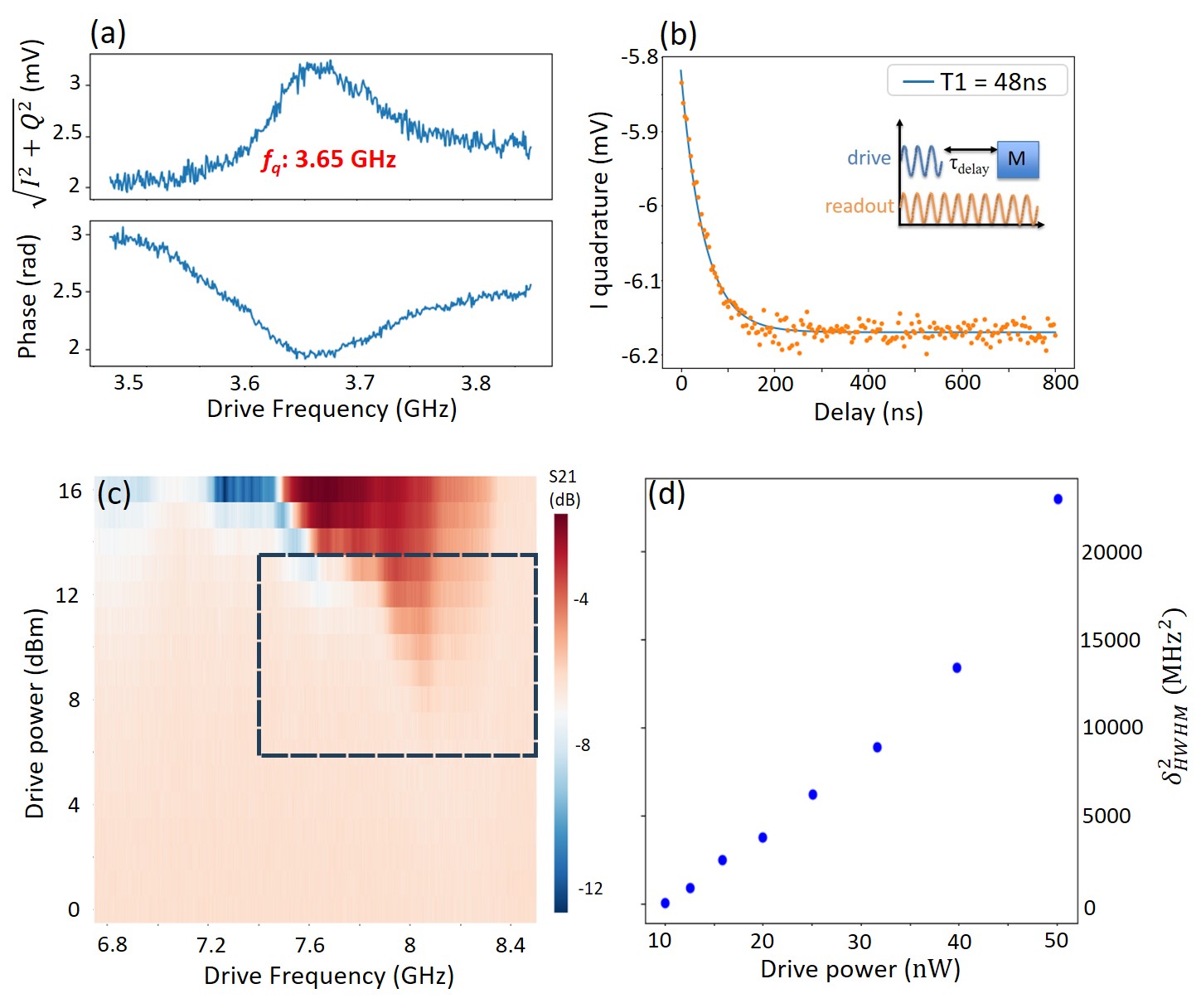}
\caption{$T_1$ measurements and and $T_2^\ast$ estimations for device 1. (a) Two-tone measurements at $\Phi = 0$ showing the amplitude (top panel) and phase (bottom panel) of $S_{21}$ as a function of the drive frequency. The $\left|0\right\rangle$-$\left|1\right\rangle$ qubit transition frequency is found to be 3.65 GHz, which we used as the drive frequency for the subsequent $T_1$ measurements. (b) $T_1$ measurements using the overlapping pulses sequence as shown in the inset. $T_1$ is extracted to be approximately 48 ns by fitting the data (solid line) to $Ae^{-\tau_{\mathrm{delay}}/T_1}$. (c) Two-tone measurement as a function of drive power and drive frequency (power-broadening measurement) performed at $\Phi = 0$ (sweet spot). (d) The squared spectral linewidth as a function of drive power, extracted from the region highlighted by the dashed square in (c). Note that (a) and (b) were obtained during the fifth cooldown, whereas (c) and (d) were obtained during the first cooldown of the device.}      
\label{FigS4}
\end{figure}

We were constrained by an equipment shortage and were only able to perform time-domain pulse measurements on device 1 during its fifth cooldown, when it was mounted in a 5.5 GHz cavity. The measurement scheme used a reflection setup as shown in Fig. \ref{FigS2} (b), and the calibrated $Q_{ext}$ values for the readout and drive ports of the 3D cavity were 928 and 4840, respectively. The qubit frequency in this cooldown was further reduced compared to cooldown 4, measuring $f_q \approx$ 3.65 GHz at $\Phi = 0$, as shown in the measured $S_{21}$ in Fig. \ref{FigS4}(a). We did not manage to measure the relaxation time $T_1$ of the qubit using the standard scheme. Instead, we followed Ref. \cite{AlbertHertel2022,Aparicio2025} and used overlapping 200 ns-long readout and drive pulses. In this scheme, the drive and readout pulses are sent at the same time, and the demodulation sequence (box labeled M) starts after a delay $\tau_{\mathrm{delay}}$, as illustrated in the inset of Fig. \ref{FigS4}(b). We extract a relaxation time $T_1$ $\approx$ 48 ns, as shown in Fig. \ref{FigS4}(b), consistent with values reported for graphene gatemons \cite{Wang2019}.

In the main text, we presented the spectral linewidth as a function of magnetic flux and estimated a lower bound for $T_2^\ast$ between 4.927 ns and 0.428 ns within the probed flux range, as shown in Fig. \ref{Fig4}(c) and (d). To further investigate this lower bound, we performed a power-broadening measurement, in which the drive frequency is swept at increasing drive power at $\Phi = 0$ (the sweet spot), as shown in Fig. \ref{FigS4}(c). The spectrum shows the linewidth of the qubit $\left|0\right\rangle$-$\left|1\right\rangle$ transition increases with enhanced drive power. The extracted spectral linewidth in the low–drive-power region, highlighted by the dashed rectangle in Fig. \ref{FigS4}(c), is squared and plotted in Fig. \ref{FigS4}(d). If $T_1$ is known, then in principle the relation $(2\pi\delta_{HWHM})^2$ = $(\frac{1}{T^\ast_2})^2+n_s\omega_{vac}^2\frac{T_1}{T^\ast_2}$ can be used to fit and extract $T_2^\ast$. However, because $T_1$ was obtained in the fifth cooldown, whereas the spectrum shown in Fig. \ref{FigS4}(c) was measured in the first cooldown, it is not appropriate to use a $T_1$
value from a different cooldown in the formula. Nevertheless, the narrowest linewidth extracted at the lowest drive power in Fig. \ref{FigS4}(d) is 18.05 MHz, which corresponds to a lower bound of $T_2^\ast$ $\approx$ 17.63 ns at the sweet spot.

\section{V. Simulations for the fixed qubit in device 2}

\begin{figure}[!t]	
\includegraphics[scale=0.8]{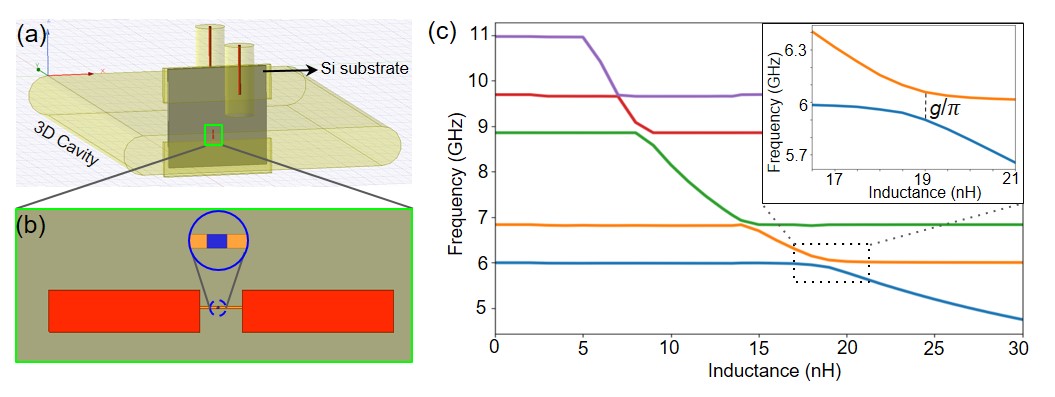}
\caption{Coupling constant and capacitance simulation of the fixed-frequency qubit in device 2. (a) 3D cavity consisting of a transmon chip. (b) Fixed frequency transmon consisting of two large capacitor pads (red boxes). The inset shows the enlarged view of the JJ (blue box). (c) Eigen-mode simulation of first five modes of the coupled qubit-cavity system, with the inset highlighting the avoided crossing between the first two modes.}      
\label{FigS7}
\end{figure}

In this work, our devices incorporate both SQUID and single-JJ circuits. For the SQUID circuits, we employed the same design as in our previous work, where the simulated capacitance and qubit–cavity coupling strength were 92 fF and $g/2\pi = 98$ MHz, respectively \cite{Chiu2025}. In this section, we present only the simulations of the single-JJ circuit (the fixed qubit), performed using Ansys HFSS. Figure \ref{FigS7} (a) shows the fixed qubit placed on a silicon substrate inside a 6 GHz cavity, with the cavity boundary assigned as a finite conductor. In Fig. \ref{FigS7}(b), the qubit comprises two large capacitor pads (red boxes) connected by a single JJ, with an enlarged view of the junction (blue box) provided in the inset. To simplify the computation, both the capacitor pads and JJ are modeled as 2D sheets, where the pads are assigned perfect conductor boundaries and the JJ is represented as a lumped RLC element with parallel inductance. By varying the junction inductance, the qubit frequency can be tuned. In a coupled system, as the qubit frequency approaches the cavity frequency, there will be deviation in cavity frequency resulting in an avoided crossing of qubit and cavity resonant frequencies. Fig. \ref{FigS7} (c) illustrates the evolution of first five eigenmodes as a function of the junction inductance. The first mode (blue curve) corresponds to the cavity frequency, while the second mode (orange curve) represents the qubit frequency. The inset shows the avoided crossing between the first two modes around an inductance of 19 nH, where the qubit and cavity frequencies coincide. From this data, the qubit-cavity coupling strength is extracted as $g/2\pi$ = 78.9 MHz. Additionally, the intrinsic capacitance between the transmon capacitor pads was simulated using Ansys Maxwell 3D. By applying the excitation voltage of 1 V to one pad and 0 V to the other pad, the inter-pad capacitance was determined to be 32.9 fF.

\section{VI. Device 2 loaded in different cavities in multiple cooldowns}

\begin{figure}[!t]	
\includegraphics[scale=0.5]{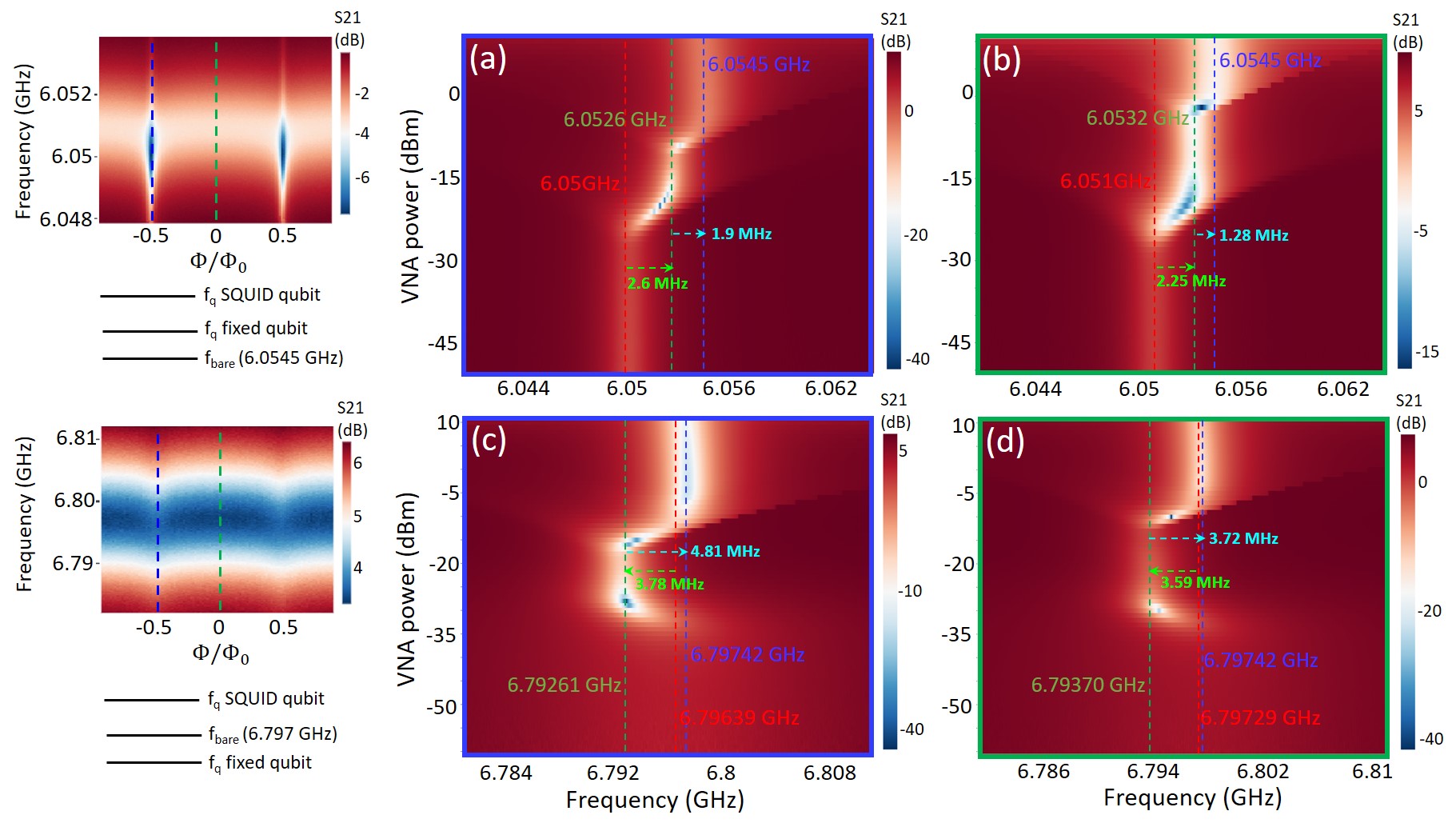}
\caption{Power dependence of device 2 at flux points corresponding to the maximum and minimum values of $f_q$ in two different cavities. (a) Power dependence of device 2 in a 6 GHz cavity, measured at the flux point indicated by the blue dashed line (where $f_q$ is minimal) in the upper-left inset. (b) Same as (a) but measured at the flux point indicated by the green dashed line (where $f_q$ is maximal) in the upper-left inset. (c) Power dependence of device 2 in a 6.8 GHz cavity, measured at the flux point indicated by the blue dashed line (where $f_q$ is minimal) in the upper-left inset. (d) Same as (c) but measured at the flux point indicated by the green dashed line (where $f_q$ is maximal) in the upper-left inset. The insets show the flux-tuning data for device 2 in a 6 GHz cavity (top) and a 6.8 GHz cavity (bottom), together with the corresponding energy detuning schemes inferred from the power-dependence measurements. In panels (a)–(d), the green dashed arrows indicate the first dispersive shift, whereas the light blue dashed arrows mark the second dispersive shift. Panels (a) and (b) were measured in cooldown 1, while (c) and (d) were measured in cooldown 2.}      
\label{FigS5}
\end{figure}

We loaded device 2 in 6 GHz and 6.8 GHz cavities and performed reflection or transmission measurements across four cooldowns (see Fig. \ref{Fig5}). For device 2 in the 6 GHz cavity during its first cooldown, we measure the power dependence at two flux points (indicated by the green and blue dashed lines in the upper-left inset of Fig. \ref{FigS5}(a)), corresponding to the maximum and minimum values of $f_q$, as shown in Figs. \ref{FigS5} (a) and (b). Fig. \ref{FigS5} (c) and (d) present the corresponding measurements for device 2 in the 6.8 GHz cavity during its second cooldown. From the flux modulation data (Fig. \ref{Fig5}) discussed in the main text, the first dispersive shift is attributed to the fixed qubit, while the second is attributed to the SQUID-based qubit. For the 6 GHz cavity, both dispersive shifts occur toward higher frequencies, suggesting that the two qubits have transition frequencies higher than the bare cavity frequency, as illustrated in the lower-left inset of Fig. \ref{FigS5} (a). In contrast, for the 6.8 GHz cavity, the first dispersive shift occurs toward lower frequency, while the second shift is toward higher frequency, suggesting a scenario as illustrated in the lower-left inset of Fig. \ref{FigS5}(c). As mentioned in the main text, the readout power required for the cavity to enter the high-power regime is characterized by the critical photon number $\overline{n}_{crit}$ $\approx$ $\frac{\Delta^2}{4g^2}$ \cite{Blais2004a,Gambetta2006,Boissonneault2008}. Comparing Fig. \ref{FigS5}(a) and (b), one can see that the onset power for the first dispersive shift is similar ($\approx$ -22.5 dBm), as the detuning $\Delta$ between the fixed qubit and the cavity is the same in both flux points. In contrast, the onset power for the second dispersive shift is higher in Fig. \ref{FigS5}(b) ($\approx$ -2.5 dBm) than in Fig. \ref{FigS5}(a) ($\approx$ -9 dBm), which can be attributed to a larger detuning $\Delta$ between the SQUID-based qubit and the cavity at the sweet spot. The same situation also applies to the 6.8 GHz cavity case, as shown in Fig. \ref{FigS5}(c) and (d), where the onset power for the first dispersive shift is around -34 dBm in both panels, while that for the second shift is higher, approximately -12 dBm in Fig. \ref{FigS5}(d) compared to -16.6 dBm in Fig. \ref{FigS5}(c). These observations support assigning the first dispersive shift to the fixed qubit and the second to the SQUID-based qubit.

We extract the dispersive shifts from the power-dependence measurements at $\Phi = 0$ (the sweet spot), as shown in the bottom panels of Fig. \ref{Fig5}. The first dispersive shift is 2.25 MHz, -3.59 MHz, -1.6 MHz and -3.32 MHz, while the second dispersive shift is 1.28 MHz, 3.72 MHz, 3.4 MHz and 4.6 MHz in cooldown 1, 2, 3 and 4. Using the second dispersive shift and adopting the coupling strength of $g/2\pi \approx 100$ MHz for the SQUID-based qubit, as discussed in the main text, we infer $f_{q,\mathrm{max}}$ to be 13.86 GHz, 9.48 GHz, 9.00 GHz, and 8.23 GHz for cooldowns 1, 2, 3, and 4, respectively. Note that we also observed a monotonic decrease in $f_{q,\mathrm{max}}$ across cooldowns, as was observed in device 1. For the fixed qubit, based on the simulated coupling strength of $g/2\pi \approx 78.9$ MHz (see section V of SM), together with the first dispersive shift extracted from the bottom panels of Fig. \ref{Fig5}, we infer the qubit frequencies $f_q$ to be 8.82 GHz, 5.059 GHz, 2.171 GHz, and 4.174 GHz for cooldowns 1 to 4, respectively. Note that the coupling strength $g$ may vary across cooldowns, as observed in device 1, which may account for the particularly low value extracted in cooldown 3.

The onset power for the SQUID-based qubit to enter the high-power regime is higher than that of the fixed qubit in all cooldowns, as seen in Fig. \ref{FigS5} and in the bottom panels of Fig. \ref{Fig5}. Using $g/2\pi$ = 78.9 MHz for the fixed qubit and $g/2\pi$ = 100 MHz for the SQUID-based qubit, together with the qubit frequencies extracted in the previous paragraph, we estimate the ratio of the critical photon number $\overline{n}_{crit}$ between the SQUID-based qubit and the fixed qubit to be approximately 4.846, 1.449, 0.344, and 0.8136 for cooldowns 1-4, respectively. The ratios obtained in cooldowns 1 and 2 agree with the experimental observations (that is, the SQUID-based qubit has a larger critical photon number and therefore a higher onset power) whereas those for cooldowns 3 and 4 do not. While the actual coupling strength $g/2\pi$ may vary across cooldowns and influence the accurate ratio estimated above, the difference in Josephson energy can account for the higher onset power observed for the SQUID-based qubit. In device 2, the junction width of the SQUID is 3.5 $\mu$m, whereas that of the single-JJ circuit is 1.5 $\mu$m [see Fig. \ref{FigS1}(g)], resulting in a larger $I_C$ (and thus $E_J$) for the SQUID compared to the single JJ. Transmons with smaller $E_J$ possess a shallower potential well and are therefore more susceptible to multi-photon excitations into higher energy levels, reaching the onset of the high-power, or “ionized,” regime at comparatively lower drive powers \cite{Koch2007,Krantz2020}. In contrast, weakly anharmonic transmons with large $E_J$ exhibit a deeper and more harmonic potential, thus requiring substantially higher input powers to enter this regime. 

\begin{figure}[!t]	
\includegraphics[scale=0.65]{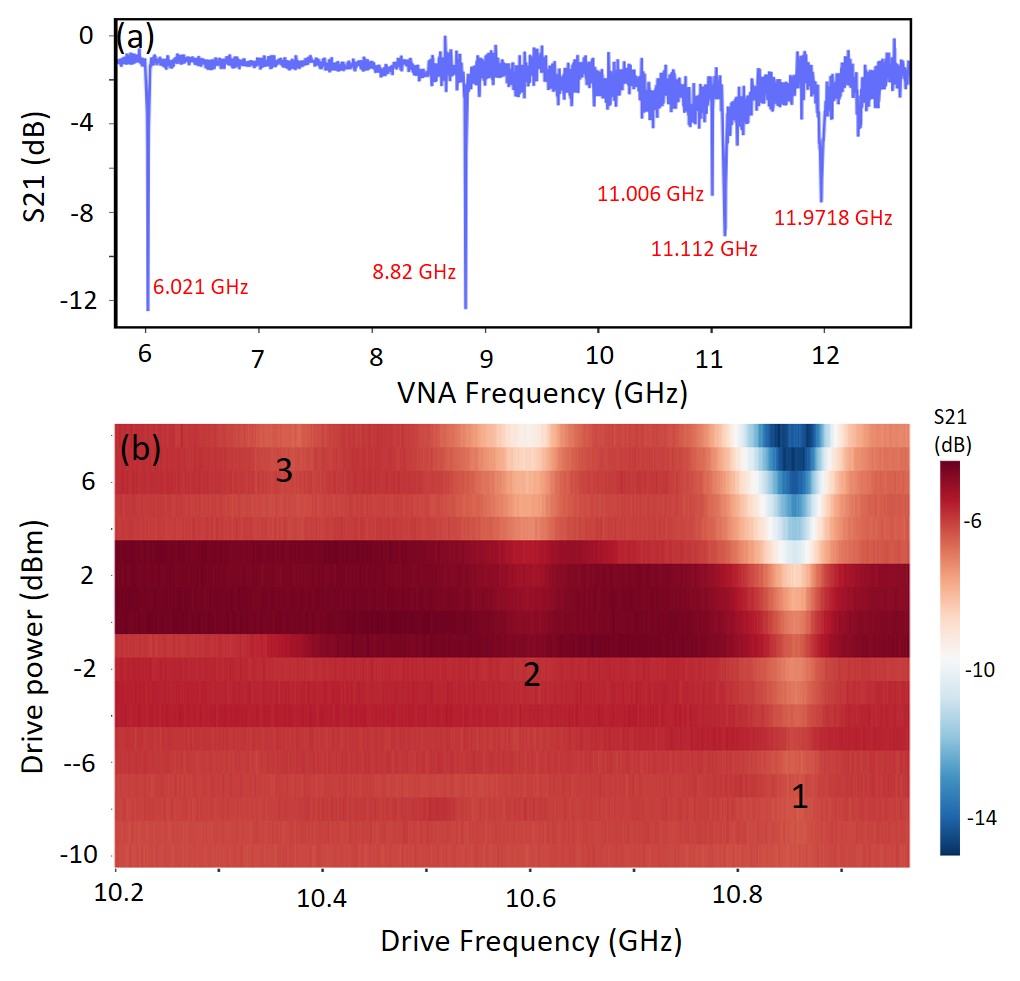}
\caption{Potential qubit transitions detected for the fixed qubit in the two-tone measurements. (a) Room-temperature $S_{21}$ measurements of the 6 GHz cavity (loaded with device 2) before cooldown 1. The TE$_{101}$ mode appears at 6.021 GHz, along with other higher-frequency modes. (b) Two-tone measurements of device 2 in cooldown 1, obtained by sweeping the drive frequency at increasing drive power, reveal three transitions labeled 1, 2, and 3. These three transitions are not matching any of the higher-frequency cavity modes shown in (a).}      
\label{FigS6}
\end{figure}

We performed continuous-wave two-tone measurements on device 2 across cooldowns; however, no flux-tunable qubit transition was detected up to a drive frequency of 12 GHz (the limit of our equipment). For the fixed qubit, because its transition frequency is not flux-tunable, we must carefully distinguish any possible qubit transition from the existing cavity modes. Fig. \ref{FigS6}(a) shows the $S_{21}$ measurements performed at room temperature, with the TE$_{101}$ mode appearing at 6.021 GHz, along with other higher-frequency modes. We did this at room temperature as the low-temperature HEMT amplifier has an operation range between 4 and 8 GHz, which prevent us from testing the higher cavity modes at low temperature. The two-tone measurements in Fig. \ref{FigS6}(b) show the possible qubit transitions detected during cooldown 1, which can be distinguished from the higher cavity modes shown in Fig. \ref{FigS6}(a). The three transitions labeled 1-3 are located at 10.86 GHz, 10.59 GHz and 10.36 GHz, which we attribute to the $f_{01}$, $f_{02}$/2 and $f_{12}$ transitions, respectively (the subscripts denote the qubit level indices). If these three transitions are associated with the fixed qubit, the separation between transition 1 and 3 is 500 MHz, which is close to the charging energy $E_C$ = $\frac{e^2}{2C}$ $\approx$ 588 MHz estimated from the simulated capacitance 32.9 fF (see section V of the SM). Using the first dispersive shift $\chi/2\pi$ $\approx$ 2.25 MHz extracted from cooldown 1 and $\Delta/2\pi = (f_{q} - f_{bare})$ = 10.86 GHz $-$ 6.0545 GHz = 4.8055 GHz, we can estimate the coupling strength to be $g/2\pi$ $\approx$ 104 MHz using $\chi = g^{2}/\Delta$. This value lies within the possible range of our simulated coupling strength, $g/2\pi$ = 78.9 MHz for the fixed qubit (see section V of the SM), given that we also observed variations in $g$ in device 1. In addition, the JJ width of the fixed qubit in device 2 is 1.5 $\mu$m, whereas the total JJ width in the SQUID of device 1 is 2 $\mu$m [see Fig. \ref{FigS1}(g)]. This implies that $E_J$ for the fixed qubit in device 2 is approximately 0.75 times that of the SQUID qubit in device 1. For the charging energy, $E_C$ of the fixed qubit in device 2 is about 2.79 times larger than that of the SQUID qubit in device 1, based on the simulated capacitances of 32.9 fF and 92 fF, respectively. Since $f_q \approx \sqrt{8E_JE_C}/h$, we expect the qubit frequency of the fixed qubit in device 2 to be $\sqrt{0.75\times2.79}$ times that of the SQUID qubit in device 1. This gives $\sqrt{2.0925} \times$ 8.068 GHz $\approx$ 11.66 GHz, which is close to the observed 10.86 GHz for transition 1. Note that we use only the data from the first cooldown of both devices for our analysis, as the devices degrade over time, leading to reduced qubit frequencies in later cooldowns. Although the above analysis suggests that these three transitions can be associated with the fixed qubit, time-domain measurements are still required to carefully examine their coherence properties. Unfortunately, due to equipment shortage, we were unable to perform further investigations during that time.

\end{document}